\documentclass[aps,prd,twocolumn,groupedaddress,showpacs,showkeys]{revtex4-1}

\usepackage{graphicx}

\begin{document}

\title{Nonlinear effects and the behavior of total hadronic and photonic cross sections}

\author{A. V. Giannini}
\email[]{avgiannini@usp.br}
\affiliation{Instituto de F\'{\i}sica, Universidade de S\~{a}o
Paulo, CEP 05315-970, S\~ao Paulo, Brazil}

\author{F. O. Dur\~aes}
\email[]{fduraes@if.usp.br}
\email[]{duraes@mackenzie.br}
\affiliation{Curso de F\'{\i}sica, Escola de Engenharia,
Universidade Presbiteriana Mackenzie, CEP 01302-907, S\~{a}o Paulo, Brazil }

\date{\today}

\begin{abstract}
In this paper we use an eikonalized minijet model where the effects
of the first nonlinear corrections to the DGLAP equations are
taken into account. The contributions coming from gluon
recombination effects are included in the DGLAP+GLRMQ approach for
the free proton in the context of saturation models. The
parameters of the model are fixed to fit total $pp$ and $\bar p p$
cross sections, including the very recent data from LHC, HiRes, and
Pierre Auger collaborations. Glauber and multiple scattering
approximations are then used to describe the inclusive inelastic
proton-Air cross section. Photoproduction cross sections, without
change of parameters fixed before, are also obtained from the
model using vector meson dominance and the additive quark model.
We show and discuss our main results as well as the implications
of saturation effects in the behavior of total hadronic and
photonic cross sections at very high energies.
\end{abstract}


\maketitle

\section{Introduction}
The growth of total hadronic cross sections with energy has been
studied for decades and many phenomenological and theoretical
efforts have been made to explain it.

One of the most important explanations for this behavior was
proposed in the 1970s \cite{cline} based on quantum chromodynamics
(QCD): the behavior of hadronic cross sections with the center of
mass energy ($\sqrt s$) was very similar to the production of
jets, indicating that partons would be playing a role in these
interactions. This basic idea has led to an approach, called
minijet model, which takes into account that the total hadronic
cross sections can be decomposed as $\sigma_0 + \sigma_{pQCD}$,
where $\sigma_0$ characterizes a nonperturbative contribution to
the process (generally taken as energy-independent at high
energies) and $\sigma_{pQCD}$ represents the semihard
contributions, calculated in perturbative QCD (pQCD) with use of
an arbitrary cutoff at low transverse momenta $p_{T_{min}}\,(>
\Lambda_{QCD})$ \cite{halzen}.

This simple model, however, violates the unitarity of the
$S$ matrix for these processes and, consequently, the Froissart
bound, which states that total hadronic cross sections cannot grow
faster than $\ln^2 (s)$ as $s \rightarrow \infty$
\cite{froissart}. At high energies, the perturbative component of
this model is dominated by gluons with very small (Bjorken-$x$)
fractional momentum and pQCD calculations, based on linear QCD
evolution (equations developed by Dokshitzer-Gribov-Lipatov-Altarelli-Parisi
(DGLAP)~\cite{dglap} and Balitsky-Fadin-Kuraev-Lipatov (BFKL)~\cite{lipatov}),
show that the minijet cross-section grows very rapidly, dictated
by a power-like energy behavior.

To explain the experimental data quantitatively, this idea was
reformulated based on the eikonal representation to ensure
unitarity and to contain this strong level of growth. Since then
the original eikonalized minijet models~\cite{durandpi} have
been revisited and several models have been proposed based upon
then \cite{wang.eiconal}. Nevertheless, many questions about the
dynamics of interaction between partons in the high energies
regime still remain open, although, in general, good descriptions
of the experimental data have been obtained with these models.

Parallel to these developments, significant progress was achieved
in theoretical physics of small $x$ and the results from HERA,
with kinematical ranges extended upwards in $Q^2$ (the
four-momentum transfer to the proton) and also downwards in $x$,
changed our view of the structure of the proton \cite{HERA.reviews}.
Deep inelastic scattering experiments showed a rapid increase in the
density of gluons as $x$ decreases and reinforced the hypothesis that
this growth could be related to nonlinear effects in gluon evolution
equations. One now knows that for values $x \leq 10^{-4}$, gluons dominate
the hadron wave functions but, of course, it is expected that the growth
of gluon densities ``saturates" at a given time.

The understanding of this expectation is related to the momentum
transfer, $\emph{\textbf{k}}_{\perp}$, and, therefore, to the
transverse size of a gluon ($\propto 1/\emph{\textbf{k}}_{\perp}$)
in semihard interactions. For large momentum transfer, the BFKL
evolution predicts a large number of small size gluons per unit of
rapidity produced through $g \rightarrow gg$ interactions. For
small momentum transfer, on the other hand, the produced gluons
overlap themselves in the transverse area and fusion processes,
$gg \rightarrow g$, also become important.

This simple scenario shows that a typical scale, $Q_{s}$, called
``saturation scale", tells us that these latter processes are small
for $\emph{\textbf{k}}^2_{\perp}>Q^2_{s}$. For low enough momentum
transfer, $\emph{\textbf{k}}^2_{\perp}<Q^2_{s}$, however, $Q_{s}$
tell us that the recombination of gluons (fusion processes) cannot
be neglected because the gluon density is large and grows with
lowering $x$. At very high energies, smaller and smaller values of
Bjorken $x$ can be accessed and, under these conditions, the
recombination mechanism becomes more and more effective resulting
in a decrease in the population of gluons and, therefore, in the
idea of ``saturation" of partonic distributions mentioned above.

Many studies were made in the latest two decades exploring this
subject and, currently, one believes that an effective theory, the
color glass condensate~\cite{cgc,CGC.theory,bk}, correctly
describes the behavior of very small-$x$ gluons in hadronic wave
functions by an infinite hierarchy of coupled evolution equations
for the correlators of Wilson lines.

According to this picture, the (highly dense) system formed at
these extreme conditions is characterized by the limitation on the
maximum phase-space parton density that can be reached in the
hadron wave function and an $x$- or energy-dependent momentum
scale, $Q_{s}(x)$, which separates dense and dilute regimes. In
the low density regime the formalism reproduces the BFKL dynamics
for partons with transverse momentum much larger than this
``saturation momentum". On the opposite side, the saturation scale
becomes large, $Q_s (x) \gg \Lambda_{QCD}$, and the formalism
predicts that partons saturate in the hadron wave function with
occupation numbers of order $1/\alpha_s(Q_s)$. In this case the
coupling constant becomes weak [$\alpha_s(Q_s) \ll 1$] and the
high energy limit of QCD can be studied using weak coupling
techniques \cite{CGC.Blaizot,Blois.Workshop.2009}.

In this work we study the influence of the gluon recombination process
assuming that such a system may be formed in hadronic and photonic
collisions at very high energies. As mentioned above, at very
small $x$, the parton distribution functions are governed by BFKL
dynamics and this mechanism leads to nonlinear power corrections
to the DGLAP evolution equations. We also adopt here the first
nonlinear (GLRMQ) terms calculated by Gribov, Levin and Ryskin
\cite{glr}, and after by Mueller and Qiu \cite{Mueller.Qiu}, to
describe experimental cross sections and make predictions.

In what follows we briefly present the standard formulations of
the eikonalized minijet model for hadronic and photonic cross
sections, the inelastic proton-nucleus cross section in the
Glauber formalism and describe the main ingredients used in our
calculations. Then we present the strategy used to fix the
parameters of the model, show and discuss our main results and, in
the last section, we outline our conclusions.

\section{Eikonalized minijet model with saturation effects}

One of the most important contributions to predict the behavior of
hadronic cross sections with the energy from the QCD parton model
was proposed by Durand and Pi \cite{durandpi} in the late 1980s
using a formalism consistent with unitarity constraints. Many
QCD-inspired models used today have their origins based on this
``eikonal" formulation, which provides a framework where the
minijet cross sections are unitarized via multiple scattering.

In this work we have used a unitarized version of the minijet
model~\cite{wang.eiconal} where the total, elastic, and inelastic
$pp (\bar p)$ cross sections are given by
\begin{eqnarray}
\sigma^{pp(\bar p)}_{tot} (s)&=& 2 \int d^2 \vec{b}\,
\{1-e^{-Im\,\chi (b,s)} cos [Re\,\chi (b,s)]\}\,,\nonumber \\
\sigma^{pp(\bar p)}_{el} (s)&=& \int d^2 \vec{b} \,|1-e^{i\,\chi
(b,s)}|^2\,,\nonumber \\
\sigma^{pp(\bar p)}_{inel} (s)&=& \int d^2 \vec{b}
\,[1-e^{-2\,Im\,\chi (b,s)}].
\label{sigpptot}
\end{eqnarray}

The eikonal function $\chi (b,s)$ in the above expressions
contains the energy and the transverse momentum dependence of
matter distribution in the colliding particles and, through the
impact parameter distribution in the $b$ space, it is given by
$\chi (b,s) = {\text Re}\, [\chi (b,s)] + i\,{\text Im}\, [\chi (b,s)]$.

The real part of $\chi (b,s)$ represents only about $4\%$ in the
ratio of the real to the imaginary part of the forward elastic
amplitude for $pp(\bar p)$ processes and therefore, as a first
approximation, we assume ${\text Re}\,\chi (b,s)=0$ in this work. We also
assume that multiple partonic interactions are Poisson distributed
with an average number separated in soft and hard processes in a
given inelastic collision, $n(b,s) \equiv 2\,{\text Im}\,\chi (b,s) =
n_{soft}(b,s) + n_{hard}(b,s) $, and can be factorized in $b$ and
$s$ as~\cite{wang.eiconal}
\begin{eqnarray}
n(b,s) &=& W(b,\mu_{soft})\,\sigma^{soft}(s)\nonumber \\
&+&\sum_{k,l}\,W(b,\mu_{hard})\,\sigma^{hard}_{kl}(s),
\label{numbcoll}
\end{eqnarray}
where $W(b,\mu_{soft})$ and $W(b,\mu_{hard})$, which represent the
effective overlap functions of the nucleons at impact parameter
$b$, are related to the nucleon form factor in hadronic and
partonic levels [normalized such that $\int{W(b,\mu)\,d^{2}
\vec{b}}=1$], and $\sigma^{soft}(s)$ and $\sigma^{hard}_{kl}(s)$
represent the behavior of the total cross sections with energy in
soft and hard [minijet production ($mj$)] regimes in ${pp(\bar
p)}$ collisions.

At low energies the hard contribution to the eikonal function is
very small. In order to describe $pp$ and $p\bar p$ scattering at the
Intersecting Storage Rings energies we parametrize the soft contribution as~\cite{wang.eiconal}
\begin{eqnarray}
W(b,\mu_{soft})&=& \frac{\mu^2_{soft}}{96\pi}(\mu_{soft}
b)^{3}K_{3}(\mu_{soft} b), \nonumber \\
\sigma^{pp}_{soft}(E_{lab})&=&47 + \frac{46}{E^{1.39}_{lab}}, \nonumber \\
\sigma^{p\bar{p}}_{soft}(E_{lab})&=&47 +
\frac{129}{E^{0.661}_{lab}} + \frac{357}{E^{2,7}_{lab}},
\label{W.de.b}
\end{eqnarray}
where $\mu_{soft}$ is an adjustable parameter, $K_{3}$ is the
modified Bessel functions and $E_{lab}$ is the proton energy in
the laboratory system (cross sections are in ${\rm mb}$).

The hard contribution is described by the minijet production in
leading order (LO) pQCD where partons are produced back to back in
the transverse plane according to the differential cross section
\cite{minijets}:
\begin{widetext}
\begin{equation}
\frac{d\sigma^{mj}_{kl}}{dy}(s)=\kappa \int dp_T^2 \,dy_2
\sum_{{i,j}}x_1\, f_{i/h_1}(x_1,Q^2) \,x_2\,f_{j/h_2}(x_2,Q^2)\,
\frac{1}{1+\delta_{kl}}
\,\left[\delta_{fk}\,\frac{d\hat{\sigma}^{ij\rightarrow
kl}}{d\hat{t}}(\hat{t},\hat{u})+\delta_{fl}\,\frac{d\hat{\sigma}^{ij\rightarrow
kl}}{d\hat{t}}(\hat{u},\hat{t})\right]\, , \label{eq:mj.cs}
\end{equation}
\end{widetext} where $h_1$ and $h_2$ denote the colliding hadrons and
$d\hat{\sigma}^{ij\rightarrow kl}/d\hat{t}$ the subprocess cross
sections \cite{sigshat}.

The rapidities of the final state partons $k$ and $l$ are labeled
by $y\,(\equiv y_1)$ and $y_2$ and the transverse momentum of each
parton by $p_T$ ($\geq p_{T_{min}}$, the smallest transverse
momentum allowed for parton scatterings). The fractional momenta
of the colliding partons $i$ and $j$ are $x_{1,2} =
p_T/\sqrt{s}\,(e^{\pm y} + e^{\pm y_2})$, i.e., the incoming
partons are collinear with the beams. The factor $1/(1 +
\delta_{kl})$ is a statistical factor for identical particles in
the final state.

In our calculations we have assumed $\kappa=1$ and only considered
the process $gg\rightarrow gg$, $gq(\bar q)\rightarrow gq(\bar q)$
and $gg\rightarrow q\bar q$ ($q \equiv u, d, s$). We also have
parametrized the ``hard" overlap functions in impact parameter
space, $W(b,\mu_{gg})$, $W(b,\mu_{gq} \equiv \sqrt{\mu_{ qq}\mu_{
gg}})$ and $W(b,\mu_{qq})$, as Fourier transforms of a dipole form
factor [see Eq. (\ref{W.de.b})]~\cite{Block.review}. The (free)
parameters $\mu_{qq}$ and $\mu_{gg}$ represent masses which
describe the ``area" occupied by quarks and gluons, respectively,
in the colliding protons.

As discussed before, even in conventional eikonalized minijet
models the rise of the total ${pp(\bar p)}$ cross section with
energy is related to the increasing probability of perturbative
small-$x$ gluon-gluon collisions: gluon distribution functions
governed by DGLAP evolution and contributions of partons with $p_T
\geq p_{T_{min}}$ dominate the integrand of Eq. (\ref {eq:mj.cs})
increasing very fast the rise of total cross section with energy.
The numerical evaluation of this partonic contribution strongly
depends upon $p_{T_{min}}$, the chosen set of parton densities
[$f_{i,j/h_{1,2}}(x_{1,2},Q^2)$] and, of course, their evolution in
this regime.

The main ingredient of our model is the introduction of nonlinear
terms in the evolution of parton densities above. In the context of
saturation models, we adopt these corrections and make use of EHKQS
parton distribution functions \cite{GLRMQ.num.PDF,nonlinearpdf},
where the GLRMQ terms are present. This allows us to test the dynamic
responsible for the rise of total ${pp(\bar p)}$ cross sections with
energy in the presence of saturation effects.

In order to compare these two different regimes, in what follows
we shall consider GRV98(LO) \cite{gluck} and CTEQ6L \cite{cteq6l}
parton densities as references. They are governed by DGLAP
evolution equations (linear regime), where, therefore, saturation
effects are clearly absent.

At this point, we would like to call attention to the latest LHC
results \cite{DATA.LHC.pp.7, DATA.LHC.pp.8}: they have provided
very valuable information on high-energy multiparticle production,
improving our theoretical understanding of soft and semihard
parton dynamics, and showed the need for adjustments and even the
reformulation of hypotheses employed in models that propose to
establish the behavior of hadronic cross sections with energy.

It is also important to note that the recent cosmic ray data from
HiRes \cite{DATA.HiRes.pp} and Pierre Auger collaborations
\cite{DATA.P.Auger} have allowed a deeper understanding about the
nature of produced particles at very high energies and stimulated
many discussions between the accelerator and cosmic-ray
communities on common issues in these areas. The energy dependence
of total hadronic cross sections is probably the most important
question for the cosmic-ray community. The phase space regions of
relevance to the development of Air showers are not directly
accessible in the currently accelerator experiments, and, because
of that, descriptions and interpretations of the data in cosmic
rays physics at high energies depend crucially on the predictions
coming from phenomenological models \cite{LHC.and.cosmic.ray}.

For this reason it is very interesting to test the range of the
model presented in this work and to verify if it permits a
satisfactory description of the data to other processes.
Obviously, our first choice is the inelastic proton-Air cross
section.

In the Glauber multiple collision model \cite{glauber} the
inelastic proton-nucleus cross section, $\sigma_{inel}^{pA}(s)$,
can be derived in the eikonal limit (straight line trajectories of
colliding nucleons) from the corresponding inelastic
nucleon-nucleon ($NN$) cross section, $\sigma_{inel}^{NN}(s)$.

At the center-of-mass energy $\sqrt s$ and the geometry of the
$pA$ collision, it is simply determined by the impact parameter of
the reaction:
\begin{equation}
\sigma_{inel}^{pA}(s)= \int d^2\vec{b} \left[1 -
e^{-\sigma_{inel}^{NN}(s)\,T_A(b)}\right].
\label{siginel.pA.final}
\end{equation}

The usual thickness function $T_A(b)$ [$\equiv \int dz
\rho_A(b,z)$, with $\int d^2\vec{b}\,T_A(b)=A$], gives the number
of nucleons in the nucleus $A$ per unit area along the $z$
direction, separated from the center of the nucleus by an impact
parameter $b$. The function $\rho_{A}(b,z)$ represents the nuclear
density of the nucleus $A$ (with radius $R_A$) and, in what
follows, we have used \cite{HDeVries}:
\begin{equation}
\rho_A(b,z)= \rho_{0}\,\{1+exp\,[(r-R_{A})/a_{0}]\}^{-1}\,,
\label{woodsaxon}
\end{equation}
where $r=\sqrt{b^{2}+z^{2}}$, $R_{A}= 1.19 \,A^{1/3}-1.61
\,A^{-1/3}\,(fm)$, $\rho_{0}$ corresponds to the nucleon density
in the center of the nucleus $A$ (not important here due the
required normalization condition on $T_A(b)$) and $a_0$ is the so-called
``diffuseness parameter" of the Woods-Saxon profile
(\ref{woodsaxon}), assumed here as a free parameter.

Our second choice to test the model are the $\gamma{p}$ and
$\gamma\gamma$ cross sections. These processes can be derived from
the $pp$ forward scattering amplitude using vector meson dominance
(VMD) and the additive quark model with the introduction of a
probability ($P_{had}^{\gamma{p}(\gamma)}$) that the photon
interacts as a hadron (see, for example, articles produced by
Block, Pancheri, and Luna with their collaborators on that subject
\cite{wang.eiconal,Block.review}).

Assuming, in the spirit of the VMD, that at high energies the photon behaves
as a hadronic state composed by two quarks, after the substitutions
$\sigma^{s,h}\rightarrow \frac{2}{3}\sigma^{s,h}$ and
$\mu_{s,h}\rightarrow \sqrt{\frac{3}{2}\mu_{s,h}}$ in both, soft and hard
components of Eq. (\ref{numbcoll}), the $\gamma{p}$ cross section can be
written as
\begin{eqnarray}
\sigma^{\gamma{p}}_{tot} (s)&=& 2\, P_{had}^{\gamma{p}}\, \int d^2 \vec{b}
\,\{1-e^{-{\text Im}\,\chi^{\gamma{p}} (b,s)} \nonumber \\
&\times& {\text cos}\, [{\text Re}\,\chi^{\gamma{p}} (b,s)]\}\,.
\label{siggammaptot}
\end{eqnarray}

To obtain the $\gamma\gamma$ cross section we apply the same procedure
as before making the substitutions $\sigma^{s,h}\rightarrow \frac{4}{9}\sigma^{s,h}$
and $\mu_{s,h}\rightarrow \frac{3}{2}\mu_{s,h}$:
\begin{eqnarray}
\sigma^{\gamma\gamma}_{tot} (s)&=& 2\, P_{had}^{\gamma\gamma}\, \int d^2 \vec{b}
\,\{1-e^{-{\text Im}\,\chi^{\gamma\gamma} (b,s)} \nonumber \\
&\times& {\text cos}\, [{\text Re}\,\chi^{\gamma\gamma} (b,s)]\}\,,
\label{siggammagammatot}
\end{eqnarray}
where $P_{had}^{\gamma\gamma} = (P_{had}^{\gamma{p}})^{2}$.

The simplest VMD formulation only assumes the lightest vector mesons
and in this case $P_{had}^{\gamma{p}}$ is given by
$P_{had}^{\gamma{p}} = \sum_{V = \rho, \omega, \phi}{\frac{4\pi
\alpha_{em}}{f_{V}^{2}}}$, where $\alpha_{em}$ is the QED coupling
constant and $f_{V}^{2}$ is the $\gamma-V$ coupling. In this work,
however, we consider $P_{had}^{\gamma{p}}$ as a free parameter
fixed by the low energy $\gamma{p}$ data.

\section{Results and discussions}

The relevant experimental data shown in the next figures
considers only the most quoted ones in the literature and can be
found in the references \cite{Block.review,DATA.sig.pp.tot,
DATA.LHC.pp.7,DATA.LHC.pp.8,DATA.HiRes.pp,DATA.P.Auger,
DATA.sig.pp.inel,DATA.pAir,DATA.gamma.p,DATA.gamma.gamma}.

The strategy adopted in our calculations is as follows: while the
soft contribution of the model was parametrized to describe the
$pp$ and $p\bar p$ scattering at low energies, at higher energies
the hard parameters were fixed [for each set of Parton Density
Functions (PDF) used] to fit the latest LHC data~\cite{DATA.LHC.pp.7,DATA.LHC.pp.8}
and the experimental results from HiRes \cite {DATA.HiRes.pp} and Pierre Auger
(converted data) \cite{DATA.P.Auger} collaborations. Once
established the energy dependence of the total and inelastic $pp$
and $p\bar p$ cross sections (\ref{sigpptot}), the inelastic
proton-Air, the total gamma-proton, and gamma-gamma cross sections
were then obtained according to Eqs. (\ref{siginel.pA.final}),
(\ref{siggammaptot}), and (\ref{siggammagammatot}).

In all figures below, solid lines represent our results with
nonlinear evolution (EHKQS) and hard parameters fixed at
$p_{T_{min}}^{2} = 1.51\, {\rm GeV}^{2}$, $\mu_{gg}=2.00\, {\rm GeV}$ and
$\mu_{q\bar{q}} = 0.70\, {\rm GeV}$. Respectively, dashed and dotted
lines show the results in the linear regime of the model with hard
parameters fixed at $p_{T_{min}}^{2} = 2.10\, {\rm GeV}^{2}$,
$\mu_{gg}=2.03\, {\rm GeV}$ and $\mu_{q\bar{q}} = 0.73\, {\rm GeV}$ (CTEQ6L),
and $p_{T_{min}}^{2} = 1.32\, {\rm GeV}^{2}$, $\mu_{gg}=1.88\, {\rm GeV}$ and
$\mu_{q\bar{q}} = 1.00\, {\rm GeV}$ (GRV98). In all cases we have
adopted $\mu^{2}_{soft}=0.7 \, {\rm GeV}^{2}$ [see Eq. (\ref{W.de.b})].

\begin{figure*}
\includegraphics[width=10.cm]{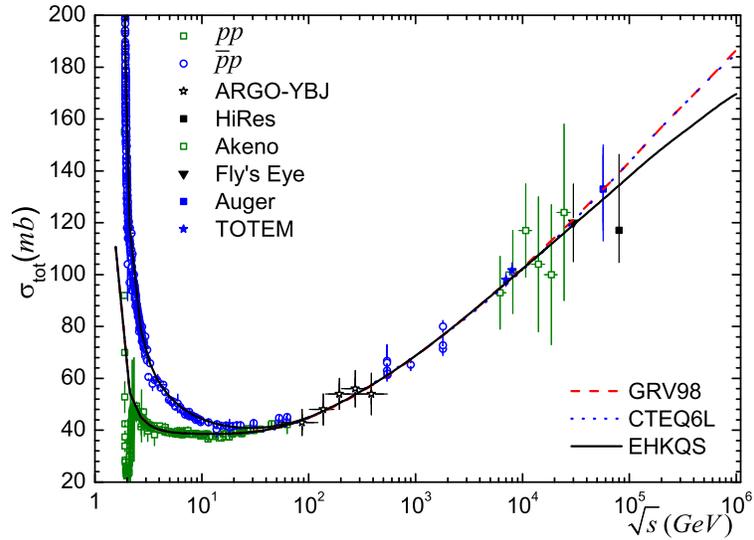}
\caption{\label{fig:fig1} Total $pp$ and $\bar p p$ cross
sections. Solid, dashed, and dotted lines represent the results of
the model [Eq. (\ref{sigpptot})] in the nonlinear and linear
regimes. The experimental data are from Refs.~\protect\cite{
DATA.sig.pp.tot,DATA.LHC.pp.7,DATA.LHC.pp.8,DATA.HiRes.pp,
DATA.P.Auger}.}
\end{figure*}

Figure \ref{fig:fig1} shows the total $pp$ and $\bar p p$ cross
sections given by Eq. (\ref{sigpptot}). The experimental data are
from Refs.~\protect\cite{DATA.sig.pp.tot,DATA.LHC.pp.7,
DATA.LHC.pp.8, DATA.HiRes.pp,DATA.P.Auger}. As can be seen, the
results obtained with nonlinear evolution also produce a good
description of the data (as good as the fits shown, for example,
in Block {\it et al.}, Pancheri {\it et al.}, and Grau {\it et al.}
\protect\cite{wang.eiconal}), in particular the most recent data
cited above (LHC, HiRes, and Pierre Auger) at higher energies.
Comparing the solid curve with the dashed and dotted ones, we
conclude that the rise of these cross sections at high energies is
still dictated by the growth of gluon densities at small $x$. The
nonlinear evolution (EHKQS), however, leads to a small softening
of these cross section due gluon fusion processes ($gg \rightarrow
g$).

\begin{figure*}
\includegraphics[width=\linewidth]{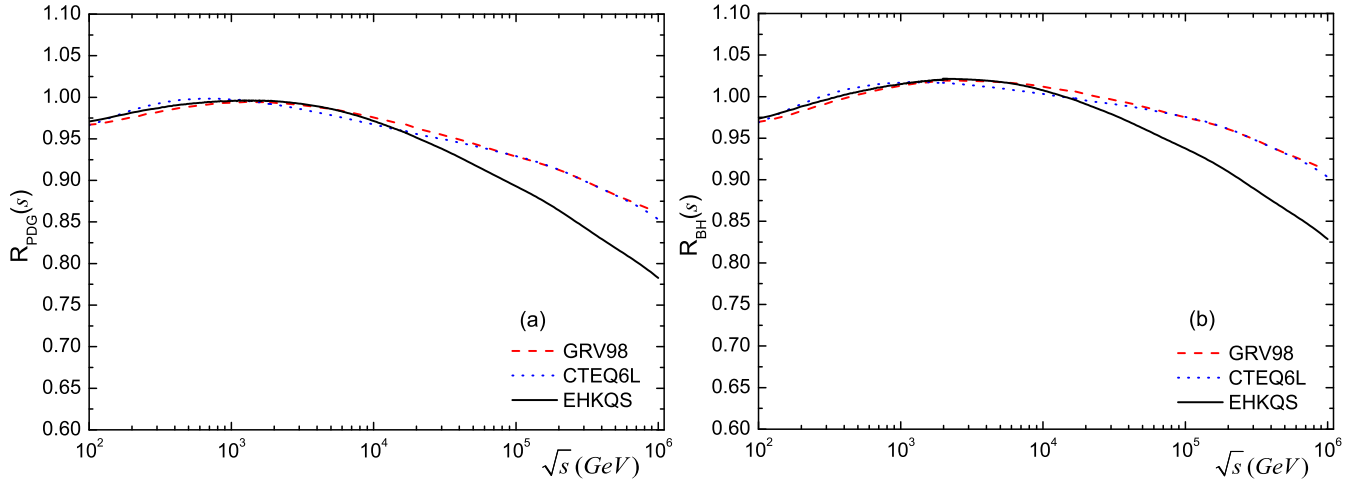}
\caption{\label{fig:fig2ab} Ratios between total cross sections
showed in the Fig. \ref{fig:fig1} and the parametrizations
given by Eqs. (\ref{PDG}) and (\ref{BH}).}
\end{figure*}

Figures \ref{fig:fig2ab} (a) and \ref{fig:fig2ab} (b) show,
respectively, the ratios between the cross sections presented in Fig.
\ref{fig:fig1} and the parametrizations from the Particle Data
Group (PDG) \protect\cite{DATA.sig.pp.tot} and from the
Block-Halzen's analysis (BH)
\protect\cite{Block.review,BlockHalzen}:
\begin{equation}
\sigma^{\mp}_{PDG}(s) = a_0\, + a_1\, {A_1}^{b_1} \mp \, a_2\,
A_1^{b_2} + a_3\, \ln^{b_3} (A_2), \label{PDG}
\end{equation}
\begin{eqnarray}
\sigma^{\pm}_{BH}(\nu) = c_0\, &+& c_1\, {C}^{d_1} \, \pm c_2\,
{C}^{d_2} \nonumber \\ &+& \, c_3 \, \ln (C)\, + c_4\, \ln^{d_3}
(C), \label{BH}
\end{eqnarray}
where the upper (lower) sign is for $pp$ ($\bar p p$) scattering,
${A_1}\equiv s/s_l$, ${A_2}\equiv s/s_h$, $C\equiv \nu/m$
[$\approx s/2m^2$, $\nu$ and $m$ represent, respectively, the
laboratory energy of the incoming proton (antiproton) and the
proton mass]. The values of the corresponding parameters are
displayed in Table \ref{tab:table1}.

\begin{center}
\begin{table}[h]
  \centering
  \caption{The fitted parameters from the quoted references through parametrizations
           (\ref{PDG}) and (\ref{BH}).}\label{tab:table1}
\begin{tabular}{|c|c|c|c|c|}
  \hline
  PDG & \protect\cite{DATA.sig.pp.tot} &   &  BH & \protect\cite{Block.review,BlockHalzen} \\
  \hline
  $a_0 \,({\rm mb})$  & $35.35  \pm 0.48 $  &   & $c_0 \,({\rm mb})$ & $37.32            $ \\
  $a_1 \,({\rm mb})$  & $42.53  \pm 1.35 $  &   & $c_1 \,({\rm mb})$ & $37.10            $ \\
  $a_2 \,({\rm mb})$  & $33.34  \pm 1.04 $  &   & $c_2 \,({\rm mb})$ & $-28.56           $ \\
  $a_3 \,({\rm mb})$  & $0.308  \pm 0.010$  &   & $c_3 \,({\rm mb})$ & $-1.440 \pm 0.070 $ \\
  $b_1$         & $-0.458 \pm 0.017$  &   & $c_4 \,({\rm mb})$ & $0.2817 \pm 0.0064$ \\
  $b_2$         & $-0.545 \pm 0.007$  &   & $d_1$        & $-0.5             $ \\
  $b_3$         & $2               $  &   & $d_2$        & $-0.585           $ \\
  $s_l ({\rm GeV}^2)$ & $1.0             $  &   & $d_3$        & $2                $ \\
  $s_h ({\rm GeV}^2)$ & $28.9 \pm 5.4    $  &   &              &                     \\
  \hline
\end{tabular}
\end{table}
\end{center}

It should be noted from these figures that, above $\sim 10\, {\rm TeV}$,
{\it all} curves fall rapidly: at $10^3\, {\rm TeV}$, the ratios with PDG
parametrization are $\sim 0.86$ for the calculations with CTEQ6L
or GRV98 and $\sim 0.79$ for EHKQS, and the ratios with BH
parametrization are $\sim 0.91$ for the calculations with CTEQ6L
or GRV98 and $\sim 0.84$ for EHKQS.

As can be seen, apparently, our results at high energies, in all
cases, seem to be not compatible with the Froissart-type behavior
contained in both parametrizations above.
However, we have checked (in $pp$ collisions, for example) that
above $\sqrt{s} \sim 6\, {\rm GeV}$, our curves GRV98, CTEQ6L and EHKQS
(dashed, dotted, and solid lines, shown in Fig. \ref{fig:fig1}),
can also be fitted by the following ``inspired" (Froissart-type)
parametrizations {\it \`a la} PDG [Eq. (\ref{PDG}), with
$s_l=s_h=1\, {\rm GeV}^2$] and {\it \`a la} BH [Eq. (\ref{BH}), with
$m=0.938\, {\rm GeV}$], respectively:
\begin{widetext}
\begin{equation}
{\widetilde{\sigma}}^{GRV98}_{CTEQ6L}(s) = (27.9\pm 0.3)\,+ a_1\,
{s}^{b_1} - \, a_2\, s^{b_2} + (0.2152\pm 0.0015)\, \ln^{2} (s),
\label{C6LGRVPDG}
\end{equation}
\begin{equation}
{\widetilde{\sigma}}^{EHKQS}(s) = (30.5\pm 0.8)\,+ a_1\, {s}^{b_1}
- \, a_2\, s^{b_2} + (0.2014\pm 0.0035)\, \ln^{2} (s),
\label{EHKQSPDG}
\end{equation}
\begin{equation}
{\widetilde{\sigma}}^{GRV98}_{CTEQ6L}(s) = (29.74\pm 0.11)\,+
49.2143\, {s}^{d_1} - \, 39.7501\, s^{d_2} - (0.252\pm
0.004)\,ln(s) + (0.2230\pm 0.0004) \ln^{2} (s), \label{C6LGRVBH}
\end{equation}
\begin{equation}
{\widetilde{\sigma}}^{EHKQS}(s) = (18.2\pm 0.5)\,+ 49.2143\,
{s}^{d_1} - \, 39.7501\, s^{d_2} + (1.80\pm 0.08)\,ln(s) +
(0.1445\pm 0.0026) \ln^{2} (s), \label{EHKQSBH}
\end{equation}
\end{widetext}
where the parameters $a_1$, $b_1$, $a_2$, $b_2$, $d_1$, and $d_2$
are the same shown in Table \ref{tab:table1}.

\begin{figure*}
\includegraphics[width=\linewidth]{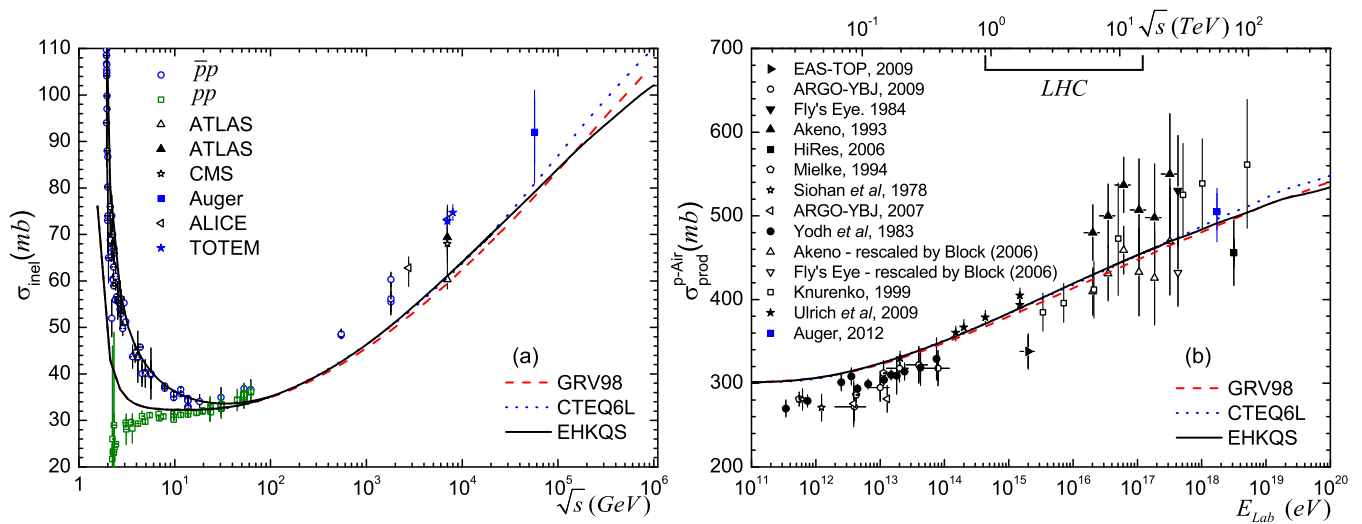}
\caption{\label{fig:fig3ab} (a) Total inelastic $pp$ and $\bar p p$
cross sections. Solid, dashed, and dotted lines represent the
results of the model [Eq. (\ref{sigpptot})] in the nonlinear and
linear regimes. The experimental data are from Refs.
\protect\cite{DATA.sig.pp.tot,DATA.LHC.pp.7, DATA.LHC.pp.8,
DATA.P.Auger,DATA.sig.pp.inel}. (b) Inelastic proton-Air cross
sections. Solid, dashed, and dotted lines represent the results of
the model [Eq. (\ref{siginel.pA.final})]. The experimental data
are from Refs.
\protect\cite{Block.review,DATA.P.Auger,DATA.pAir}.}
\end{figure*}

\begin{figure*}
\includegraphics[width=\linewidth]{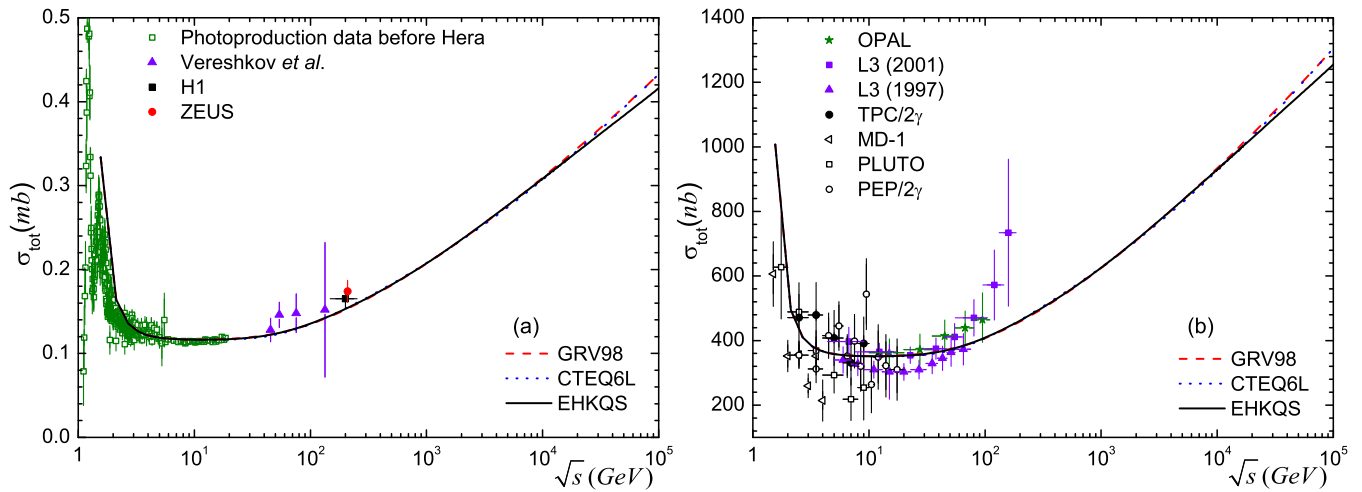}%
\caption{\label{fig:fig4ab} (a) Total gamma-proton cross sections.
Solid, dashed and dotted lines represent the results of the model
[Eq. (\ref{siggammaptot})]. The experimental data are from
Refs.~\protect\cite{DATA.sig.pp.tot,DATA.gamma.p}. (b) Total
gamma-gamma cross sections. Solid, dashed, and dotted lines
represent the results of the model [Eq. (\ref{siggammagammatot})].
The experimental data are from Ref.\protect\cite{DATA.gamma.gamma}.}
\end{figure*}

Despite that, even within the uncertainties of parameters
shown in Table \ref{tab:table1}, our parametrized results
present a very different behavior of our cross sections than those
predicted by Eqs. (\ref{PDG}) and (\ref{BH}). They suggest that it
may be much more modest with the energy, especially if nonlinear
effects are taken into account. At high energies the coefficients
of $ln(s)$ and $ln^2(s)$ in all parametrizations above are very
important and determine the growth of the hadronic cross sections
with energy. Of course, the rise of the total hadronic
cross sections at the highest energies still constitutes an open
problem \cite{Menon}, demanding further and detailed
investigations (theoretical and experimental), and also, our model
and results will be tested in the next high energies experiments.

The results for the inelastic $pp$ and $\bar p p$ cross sections
[Eq. (\ref{sigpptot})] are shown in Fig. \ref{fig:fig3ab}(a).
The experimental data include the most recent results from the Pierre
Auger collaboration \cite{DATA.P.Auger}, LHC \cite{DATA.LHC.pp.7,
DATA.LHC.pp.8,DATA.sig.pp.inel} and the oldest ones from the
Particle Data Group \cite{DATA.sig.pp.tot}, where was considered
$\sigma_{inel}^{pp(\bar p)}(s)= \sigma_{tot}^{pp(\bar p)}(s)-
\sigma_{el}^{pp(\bar p)}(s)$.

In all the cases the description of the data is not good. As
discussed above, the parameters of the present model were fixed
{\it only} through the {\it total} experimental $pp$ and $\bar p
p$ cross sections. They determine the behavior of the imaginary
part of the eikonal function with energy and do not provide a
satisfactory description of the experimental inelastic $pp$ and
$\bar p p$ cross sections. It is important to note, however, that
conventional eikonal models possess the same problem and also
cannot be able to produce a reasonable description of these data.

In order to modify this behavior, some alternative QCD-inspired
eikonal models have been proposed. For example, Luna and
collaborators \protect\cite{wang.eiconal} have introduced
{\it ad hoc} an infrared dynamical gluon mass scale in the
calculations of $pp$ and $p \bar p$ forward scattering quantities,
which (as claimed by the authors) allows one to describe successfully
the forward scattering quantities, $\sigma_{tot}^{pp(\bar p)}$,
$\rho$ (the real to the imaginary part of the forward scattering
amplitude), the ``nuclear slope" $B$ and differential cross
section $d\sigma^ {pp(\bar p)}_{el}(s,t) / dt$, in excellent
agreement with the available experimental data (at least up to
$\sqrt s = 1.8\, {\rm TeV}$). Gribov and collaborators
\protect\cite{glr}, on the other hand, have been proposed an
energy dependent cutoff at low transverse momentum, which,
effectively, mimics saturation effects. Another interesting
approach assumes a resummation of soft gluon emission (down to
zero momentum) to soften the rise of the total cross section due
to the increasing number of gluon-gluon collisions at low $x$
(Grau {\it et al.} \protect\cite{wang.eiconal}). The model
presented in this work, as can be noted, does not include any of
these approaches.

As in other eikonal models cited before, on the other hand, our
model also does not include diffractive processes. As pointed out
in \protect\cite{Lipari}, the recently published measurements of
$p p$ inelastic diffraction cross sections at LHC indicate that
the rates of diffractive events in inelastic collisions, estimated
from the pseudorapidity distributions of charged particles, are
$\sigma_{SD}/\sigma_{inel}\simeq 0.20$ for single diffraction
processes and $\sigma_{DD}/\sigma_{inel}\simeq 0.12$ for double
diffraction processes (diffractive mass $M^2_X < 200\, {\rm GeV}^2$ and
collisions at $\sqrt{s} \sim 1 - 7\, {\rm TeV}$). Of course, an inelastic
diffractive component would be desirable and necessary in our
model in order to reconcile the estimates of $\sigma^{pp}_{inel}$
with all experimental data ($\sigma_{diff}$), but its inclusion in
a consistent way would require a more complex framework than that
used here.

Albeit complemental, this figure [Fig. \ref{fig:fig3ab}(a)]
shows the role played by the saturation effects at very high
energies, and tells us that the inclusion of other approaches or
diffractive processes may be really needed.

Figure \ref{fig:fig3ab}(b) shows the results for the inelastic
proton-Air cross sections given by Eq. (\ref{siginel.pA.final}).
They take into account only the geometry of the $p-{\text Air}$ collision
[we have fixed $A\equiv A_{Air} = 14.5$ and $a_0 = 0.75\, {\rm fm}$, see
Eq. (\ref{woodsaxon})] and the semihard dynamics of the model, via
the imaginary part of the eikonal function contained in the
inelastic cross sections (determined by the parameters fixed to
describe total $pp$ and $\bar p p$ data cross sections).

As discussed above, the description of the experimental $pp$ and
$\bar p p$ inelastic cross sections is not good. Nevertheless, the
results obtained with this simple approach (even with nonlinear
evolution) at higher energies have the same quality of Monte Carlo
predictions \protect\cite{Cartiglia,Ulrich}, even though at lower
energies, of course, the description is worse than that of such
models. But it is interesting to note that our results favor a
moderately slow rise of the proton-Air cross section towards
higher energies, as indicated by the Pierre Auger measurement
\protect\cite{DATA.P.Auger}. These results will have implications
about future measurements at the LHC (whose first analysis also
indicates slightly smaller hadronic cross sections than expected
within many models) and, certainly, will be test the hypothesis,
dynamics, and predictions of the models.

Figures \ref{fig:fig4ab} (a) and \ref{fig:fig4ab} (b) show our
results for the total $\gamma p$ and $\gamma \gamma$ cross
sections, respectively given by Eqs. (\ref{siggammaptot}) and
(\ref{siggammagammatot}). The experimental data are from
Refs.~\protect\cite{DATA.sig.pp.tot,DATA.gamma.p} and
\protect\cite{DATA.gamma.gamma}. The probability that in these
collisions the photon interacts as a hadron was fixed at
$P_{had}^{\gamma{p}}=1/221$.

Considering that these cross sections are determined by the same
parameters used in $pp$ and $\bar p p$ collisions and by the
changes {\it only} coming from the ``weights" introduced by the
VMD and additive quark model, in all the cases (including in the
nonlinear regime), the agreement with data points is reasonable.

The model underestimates the latest data points for total $\gamma
p$ cross sections and cannot describe the data above $\sqrt{s}\sim
100\, {\rm GeV}$ for $\gamma \gamma$ cross sections. But, as discussed
above, this is also a problem for other models in the literature.
To circumvent this problem, for example, Luna and Natale
\protect\cite{wang.eiconal} suggest that the probability that a
photon interacts as a hadron increases logarithmically with $\sqrt
s$ [like $P_{had} = a + b \, ln(s)$]. Even then, they only get a
successful description with this hypothesis if their results are
compared to the (OPAL and L3) data, handled with the PYTHIA and
PHOJET Monte Carlo generators. The Bloch-Nordsieck formalism to
these collisions (Grau {\it et al.} \protect\cite{wang.eiconal}),
on the other hand, is obviously more robust. It basically depends
on the choice of minimum hard cutoff ($p_{T_{min}}$, related to
the chosen PDF set) and on the infrared parameter ($p$), which
controls the quenching of the (minijet cross section) rise at high
energy and, consequently, the absolute value of $n^{hard}$ [see
Eq. (\ref{numbcoll})]. Models constructed from that describe more
satisfactorily the experimental data at energies around $100\,
{\rm GeV}$ but are strongly dependent on the infrared parameter above.

\section{Conclusions}

In this work we have used an eikonalized minijet model, where the
effects of the first nonlinear corrections to the DGLAP equations
are taken into account, to investigate (simultaneously) the energy
behavior of total hadronic and photonic cross sections.

First of all, we call attention that the main dynamical ingredients
of our model are completely determined by the choices of the PDF
used and, consequently, by the hard parameters [presented in Eq.
(\ref{eq:mj.cs})], which have been fixed to fit the latest LHC
data and the experimental results from HiRes and Pierre Auger
(converted data) collaborations, for $pp$ and $p\bar p$ collisions
at high energies. The behavior with the energy of other hadronic
and photonic cross sections studied here is, therefore, dictated
by these conditions (especially on the hard parameters).

We do not have a good description of inelastic $pp$ and $p\bar p$
data cross sections in any case, which seems to indicate the needed
for different dynamical ingredients into the semihard component of
the model. Despite that, even though the procedure adopted might
be considered ``particular", the introduction of nonlinear
corrections into the model allows a satisfactory description of
experimental cross sections investigated in this work and should
not be discarded {\it a priori} for this processes. The
corrections are relatively small and only manifest themselves at
extremely high energies, but our results show that the saturation
effects attenuate more strongly the growth of total hadronic and
photonic cross sections than those obtained by conventional
eikonal models governed by the linear regime.

We also call attention that, as discussed above, our results can
be described through a Froissart-type behavior (dictated by the
PDG and BH parametrizations quoted in the literature). But, when
the nonlinear effects are included in the calculations, our EHKQS
results exhibit a very moderated behavior with energy coming from
the gluon recombination effects. This is the main result of our
study.

The fact that we have achieved reasonable success in using this
simple model encourages us to test it to try to describe another
observable where, perhaps, nonlinear effects of QCD can be more
evidently present (works are in progress).

\vspace{0.5cm} \underline{Acknowledgements}: This work has been
supported by the Brazilian funding agencies CAPES, CNPq and FAPESP.
We would like to thanks E.G.S. Luna, F.S. Navarra and V.P. Gon\c{c}alves
for several instructive discussions.

\end{document}